\documentclass[final,5p,times, sort&compress, fleqn,twocolumn]{elsarticle}
\journal{Journal of Magnetism and Magnetic Materials}
\usepackage{amsmath}
\usepackage{bm}
\usepackage{hyperref}
\hypersetup{colorlinks = true}
\usepackage{breakurl}

\begin{document}

\begin{frontmatter}

  \title{
    Distribution of write error rate of spin-transfer-torque magnetoreistive random access memory caused by a distribution of junction parameters\tnoteref{grant}
  }

  \author{Hiroshi Imamura}
  \ead{h-imamura@aist.go.jp}

  \author{Hiroko Arai}
  \ead{arai-h@aist.go.jp}

  \author{Rie Matsumoto}

  \address{National Institute of Advanced Industrial Science and Technology (AIST), Tsukuba, Ibaraki 305-8568, Japan}

  \tnotetext[grant]{This work is partly supported by JSPS KAKENHI Grant Numbers JP19H01108, No. JP20K12003.}

  \begin{abstract}
    Distribution of write error rate (WER) of spin-transfer-torque magnetoreistive random access memory (STT MRAM) caused by a distribution of resistance area product and anisotropy constant is theoretically studied.  Assuming that WER is much smaller than unity, and junction parameters obey a normal distribution, we show that the WER obeys a logarithmic normal distribution. We derive analytical expressions for the probability density function and statistical measures. We find that the coefficient of variation of WER can be reduced by decreasing the pulse width. We also perform numerical simulations based on the Fokker-Planck equation and confirm the validity of the analytical expressions. The results are useful for designing reliable STT MRAMs.
  \end{abstract}

  \begin{keyword}
    spin transfer torque, magnetoresistive random access memory, write error rate, probability distribution function, logarithmic normal distribution
  \end{keyword}

\end{frontmatter}

%========================================
% Introduction
%========================================
\section{Introduction}
Spin-transfer-torque magnetoreistive random access memory (STT MRAM) has been attracting much attention as a key component for future low-power electronics because of its useful characteristics such as high integration density, non-volatility, low-latency, and high-endurance \cite{Yuasa2013,Ando2014,Kent2015,Apalkov2016,Sbiaa2017,Cai2017,Garzon2021,Na2021,Worledge2022}. In STT MRAM information is written as stable magnetic states which are separated by energy barrier due to magnetic anisotropy by using the STT switching method \cite{Slonczewski1996,Berger1996,Sun2022}. Magnetoresistance effect is used to read the information. The magnetic tunnel junction (MTJ) which comprises a MgO insulating barrier sandwiched by Fe-based magnetic electrodes is widely used as a basic element of STT MRAM because of the large magnetoresistance ratio \cite{Parkin2004,yuasa2004,Ikeda2008} as well as of perpendicular magnetic anisotropy \cite{Nakayama2008,Ikeda2010,Meng2011}. The perpendicularly magnetized MgO-based MTJ paved the way for a variety of applications of STT MRAM \cite{Naik2019,Gallagher2019,Aggarwal2019}. For all applications reliability is an important quality factor.

Write error rate (WER), i.e. probability of switching failure, is a key metric to characterize the reliability of STT MRAM \cite{He2007,Worledge2010,Min2010,Nowak2011,Sun2012,Butler2012,Liu2014,Matsumoto2015a,Apalkov2016,Song2020}. Magnetization switching by STT is an intrinsically stochastic process because the magnetization dynamics is disturbed by thermal agitation fields. Although much effort has been devoted to the study of WER of single memory cell, little attention has been paid to the statistical properties of an ensemble of memory cells with a distribution of junction parameters such as resistance area product (RA) and anisotropy constant. For developing a reliable STT MRAM it is important to understand the impact of a distribution of junction parameters on a distribution of WER of an ensemble of memory cells and to find a way to reduce the coefficient of variation (CV), i.e. the ratio of the standard deviation to the expectation value, of WER.

Recently Arai et al. studied the probability distribution of WER of voltage controlled (VC) MRAM \cite{Arai2021}. VC MRAM is another type of MRAM which utilizes the voltage controlled magnetic anisotropy effect to switch the magnetization and is in the basic research stage \cite{Maruyama2009,Shiota2012,Kanai2016,Grezes2016,Nozaki2019}. Assuming that the anisotropy constant of memory cells obeys a normal distribution they derived an analytical expression of the probability density function (PDF) of WER and classify the shape of PDF into two classes depending on the mean and standard deviation of the anisotropy constant. Their analysis can be applied to the case of STT MRAM.

%========================================
% In this paper
%========================================
In this paper, following Ref. \cite{Arai2021}, we analyze the distribution of WER of STT MRAM caused by a distribution of RA and anisotropy constant. Assuming that WER is much smaller than unity, and the junction parameters obey a normal distribution, we show that the WER obeys a logarithmic normal distribution. Analytical expressions for the PDF and statistical measures are derived, which show that the CV of WER can be reduced by decreasing the pulse width. The validity of analytical expressions are confirmed by numerical simulations based on the Fokker-Planck (FP) equation.

\section{Theoretical model}
\begin{figure}[t]
  \centerline{ \includegraphics[width=\columnwidth]{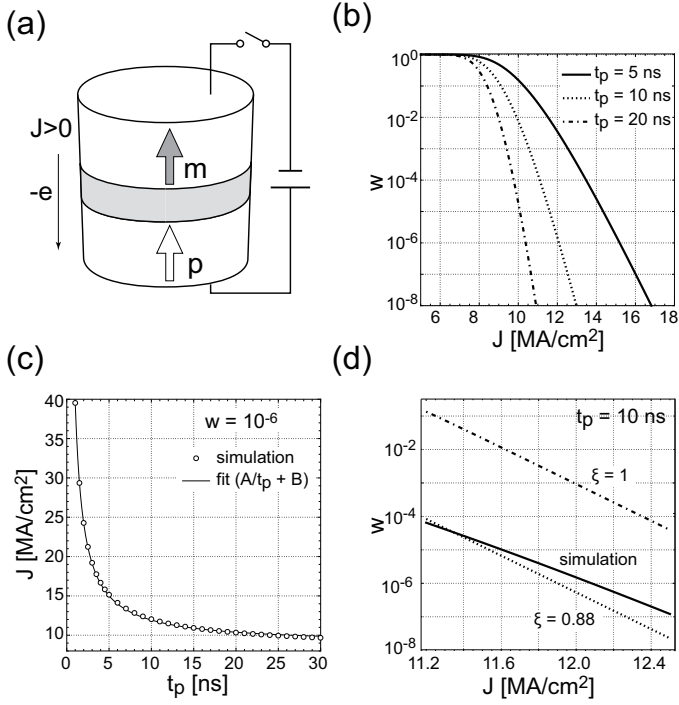}}
  \caption{
  \label{fig:fig1}
  (a) Magnetic tunnel junction nano pillar with circular cylinder shape.
  The magnetization unit vectors in the free layer (FL) and in the reference layer (RL) are represented by ${\bm m}$ and $\bm{p}$, respectively.  The positive current density, $J>0$, is defined as electrons flowing
  from the FL to the RL.
  (b) $J$ dependence of WER, $w$, for pulse width of $t_{p}$ = 5 ns (solid), 10 ns (dotted), and 20 ns (dot-dashed).
  (c) $t_{p}$ dependence of $J$ required to achieve $ w=10^{-6}$. The open circles indicate the simulation results. The solid curve represents a fit by the function of $A/t_{p} + B$, where $A$ and $B$ are fitting parameters.
  (d) $J$ dependence of $w$ for $t_{p} = 10$ ns. The solid curve indicates the simulation results. The dotted curve shows Eq. \eqref{eq:wer_approx} with $\xi=0.88$, where $\xi$ is a renormalization coefficient of anisotropy constant. The dot-dashed curve shows Eq. \eqref{eq:wer_approx} with $\xi=1$.
  }
\end{figure}

%========================================
% model
%========================================
We analyze the STT switching of the magnetization in the free layer (FL) of a circular-shaped MTJ-nanopillar shown in Fig. \ref{fig:fig1}(a). The insulating layer indicated in gray is sandwiched by the two ferromagnetic layers: the FL and the reference layer (RL). The direction of the magnetization in the FL is represented by the unit vector $\bm{m} = (m_{x}, m_{y}, m_{z})$. The magnetization unit vector in the RL is represented by $\bm{p}$ and is fixed to align in the positive $z$ direction, i.e. $\bm{p}=(0,0,1)$. The $z$ axis is taken to be the out-of-plane direction and the $x$ and $y$ axes are taken to be the in-plane directions. The positive current density, $J>0$, is defined as electrons flowing from the FL to the RL.  The size of the nanopillar is assumed to be so small that the magnetization dynamics can be described by the macrospin model.

%==============================
% Fig. 1
%==============================

The dynamics of the magnetization unit vector in the FL are calculated by solving the following Landau-Lifshitz-Gilbert (LLG) equation with STT term,
\begin{align}
  \label{eq:llg}
  \frac{d\bm{m}}{dt} = -\gamma \bm{m}\times \bm{H}_{\rm eff} -\gamma
  \chi \bm{m}\times\left(\bm{m}\times\bm{p}\right)
  +\alpha
  \bm{m}\times\frac{d\bm{m}}{dt},
\end{align}
where the first, second, and third terms on the right hand side
represent the torque due to the effective field, $\bm{H}_{\rm eff}$,
STT, and damping torque, respectively.  Here $\gamma$ is the gyromagnetic ratio, $\chi$ is the coefficient of STT, and $\alpha$ is the Gilbert damping constant.
The effective field comprises
the anisotropy field, $\bm{H}_{\rm anis}$, and the thermal
agitation field, $\bm{H}_{\rm therm}$, as
\begin{equation}
  \bm{H}_{\rm eff} =  \bm{H}_{\rm anis} +  \bm{H}_{\rm therm}.
\end{equation}
The anisotropy field is given by
\begin{equation}
  \bm{H}_{\rm anis} = \frac{2 K m_{z}}{\mu_{0}M_{s}}\bm{e}_{z},
\end{equation}
where $K$ is the anisotropy constant, $\mu_{0}$ is vacuum permeability, and $M_{s}$ is the saturation magnetization, $\bm{e}_{z}$ is the unit vector in the positive $z$ direction.
The thermal agitation field is determined by the fluctuation-dissipation theorem
\cite{Brown1963,Callen1951,Callen1952a,Callen1952b} and
satisfies the following relations: $\left\langle H_{\rm therm}^{i}(t) \right\rangle = 0$ and
\begin{align}
  \left\langle H_{\rm therm}^{i}(t)\, H_{\rm
      therm}^{j}(t')\ \right\rangle = \mu\,\delta_{i,j}\, \delta(t-t'),
\end{align}
where $\langle\ \rangle$ represents the statistical mean,
indices $i$, $j$ denote the $x$, $y$, and $z$ components of the
thermal agitation field. $\delta_{i,j}$ represents Kronecker's delta,
and $\delta(t-t')$ represents Dirac's delta function.
The coefficient $\mu$ is given by
\begin{equation}
  \mu = \frac{2\alpha k_{B} T}{\gamma\, \mu_{0}\, M_{s}\,
    \Lambda},
\end{equation}
where $k_{B}$ is the Boltzmann constant, $T$ is temperature, and
$\Lambda$ is the volume of the FL.  The coefficient of STT, $\chi$, is defined as
\begin{equation}
  \chi = \frac{\hbar P J}{2 e \mu_{0} M_{s} d},
\end{equation}
where $\hbar$ is Dirac's constant, $P$ is the spin polarization of the current, $e$ is the
elementary charge, $d$ is the thickness of the FL \cite{Slonczewski1996,Stiles2005}. The angle dependence of $\chi$ is neglected for simplicity.
At $T=0$ the critical current density over which $\bm{m}$ is switched by STT is determined by competition between the STT and the damping torque and is obtained as \cite{Sun1999,Sun2000,Lee2005}
\begin{align}
  \label{eq:Jc}
  J_{c}
  =
  \frac{4 \alpha e d  K}{\hbar P}.
\end{align}

%========================================
% Parameters
%========================================
Throughout this paper, the following typical parameters are assumed. $\alpha$ = 0.05, $K$ = 0.11 MJ/m$^{3}$, $M_{s}$ = 1 MA/m.  The diameter of the MTJ nano-pillar is 40 nm. The
thickness of the FL is $d$ = 1.1 nm. The spin polarization of
current is $P$ = 0.6, the RA is 10 $\Omega\mu$m$^{2}$, and the temperature is $T$ =300 K. These parameters give the thermal stability factor of $\Delta_{K}$ = 60 and the critical current density for STT switching of $J_{c}$ = 10 MA/cm$^{2}$.

\section{Simulation method}
\label{sec:method}
Magnetization switching is an intrinsically stochastic process because of thermal agitation. Effects of thermal agitation on STT induced magnetization switching can be analyzed based on the FP equation.  Following Brown \cite{Brown1963} we introduce the spherical coordinate defined as  $\bm{m}$ = ($\sin\theta\cos\phi$, $\sin\theta\sin\phi$, $\cos\theta$), where $\theta$ and $\phi$ are the polar angle and the azimuthal angle, respectively. The direction of $\bm{m}$ is represented by the point on a unit sphere identified by the angles $\theta$ and $\phi$. The statistical properties of $\bm{m}$ are represented by the PDF, $F(\theta,\phi)$. Since the system has a rotational symmetry around $z$ axis, the statistical properties do not depend on $\phi$. Introducing the PDF of $\theta$ defined as $W(\theta)$=$\int_{0}^{2\pi}F(\theta,\phi)d\phi$, the FP equation for $W(\theta)$ is obtained as
\begin{align}
  \label{eq:fp_zsym}
  \frac{\partial W}{\partial t}
   & =
  a
  \frac{1}{\sin\theta}
  \frac{\partial}{\partial \theta}
  \left[
    \sin\theta
    \left(\frac{\partial\epsilon}{\partial\theta}\right)
    W
    +
    \frac{1}{\beta}
    \sin\theta\frac{\partial W}{\partial \theta}
    \right],
\end{align}
where $\beta=\Lambda/(k_{B}T)$ is the inverse of the thermal energy density, the coefficient $a$ is given by
\begin{equation}
  \label{eq:coeff_b}
  a=\frac{\alpha\, \gamma}{\left(1+\alpha^{2}\right)M_{s}},
\end{equation}
and $\epsilon$ is the effective energy density defined as
\begin{equation}
  \label{eq:energy_density}
  \epsilon
  =
  K \sin^{2}\theta
  + \frac{\mu_{0}M_{s}^{2}}{\alpha}\chi\,\cos\theta.
\end{equation}
Introducing $\zeta=\cos\theta$ the FP equation and the effective energy density are expressed as
\begin{align}
  \label{eq:fp_zsym2}
  \frac{\partial W}{\partial t}
   & =
  a
  \frac{\partial}{\partial \zeta}
  \left\{
  \left(
  1-\zeta^{2}
  \right)
  \left[
    \left(
    \frac{\partial\epsilon}{\partial\zeta}
    \right)
    W
    +
    \frac{1}{\beta}
    \left(
    \frac{\partial W}{\partial \zeta}
    \right)
    \right]
  \right\},
\end{align}
and
\begin{align}
  \epsilon
  =
  K\left(1-\zeta^{2}\right)
  +
  \frac{\mu_{0}M_{s}^{2}}{\alpha}\chi \zeta.
\end{align}
Then we introduce the dimensionless time, $\tau$, thermal stability factor, $\Delta_{K}$, and the dimensionless parameter characteristic for STT, $\Delta_{J}$, which are respectively defined as
\begin{align}
  \tau
  =
  \frac{\alpha\gamma}{1+\alpha^{2}}
  \frac{k_{B}T}{M_{s}\Lambda}
  t,
\end{align}

\begin{align}
  \label{eq:Delta_K}
  \Delta_{K}
  =
  \frac{K\Lambda}{k_{B}T},
\end{align}
and
\begin{align}
  \label{eq:Delta_J}
  \Delta_{J}
  =
  -\frac{\mu_{0}M_{s}^{2}\chi\Lambda}{\alpha k_{B}T}
\end{align}
to obtain the dimensionless form of the FP equation,
\begin{align}
  \label{eq:fp_zeta}
  \frac{\partial W}{\partial \tau}
   & =
  - 2 \Delta_{K}\,   W
  + 2 \Delta_{J}\, \zeta  W
  + 6 \Delta_{K}\, \zeta^{2}  W
  \notag \\
   &
  \hspace{1em}
  - \Delta_{J}
  \left(1-\zeta^{2} \right)
  \frac{\partial W}{\partial \zeta}
  - 2 \Delta_{K}\, \zeta
  \left(1-\zeta^{2} \right)
  \frac{\partial W}{\partial \zeta}
  \notag \\
   &
  \hspace{1em}
  +
  \frac{\partial}{\partial \zeta}
  \left[
    \left( 1-\zeta^{2} \right)
    \left(
    \frac{\partial W}{\partial \zeta}
    \right)
    \right].
\end{align}
Equation \eqref{eq:fp_zeta} is solved by using the Legendre polynomial expansion,
\begin{align}
  \label{eq:LegendrePolynomialExpansion}
  W(\tau,\zeta) = \sum_{n=0}^{\infty} c_{n}(\tau) P_{n}(\zeta),
\end{align}
where $P_n(\zeta)$ is the $n$th Legendre function. Substituting Eq. \eqref{eq:LegendrePolynomialExpansion} into Eq. \eqref{eq:fp_zeta} we obtain the following equation of motion for the coefficient of the Legendre polynomial,
\begin{align}
  \frac{\partial c_{n}(\tau)}{\partial\tau}
   & =
  2\Delta_{K}
  \frac{(n-1)n(n+1)}{(2n-3)(2n-1)}
  c_{n-2}(\tau)
  \notag          \\
   & \hspace{1em}
  +
  \Delta_{J}
  \frac{n(n+1)}{2n-1}
  c_{n-1}(\tau)
  \notag          \\
   & \hspace{1em}
  +
  n(n+1)
  \left[
    \frac{2\Delta_{K}}{(2n-1)(2n+3)} -1
    \right]
  c_{n}(\tau)
  \notag          \\
   & \hspace{1em}
  -
  \Delta_{J}
  \frac{n(n+1)}{2n+3}
  c_{n+1}(\tau)
  \notag          \\
   & \hspace{1em}
  -
  2\Delta_{K}
  \frac{n(n+1)(n+2)}{(2n+3)(2n+5)}
  c_{n+2}(\tau).
\end{align}

The initial distribution is prepared by relaxing $W(\zeta)$ from the delta function at $\zeta=1$ for 5 ns without applying current. Then switching dynamics of $W(\zeta)$ are calculated under application of current during the pulse width, $t_{p}$. After the pulse the magnetization is relaxed without applying current for 5 ns. Then the WER is evaluated by integrating $W(\zeta)$ from $0$ to $1$. The basis set with 100 Legendre functions has already been enough for a converged result. The validity of the preparation procedure of the initial distribution is discussed in \ref{sec:appendixA}.

%========================================
% Results
%========================================
\section{Results}
\subsection{WER without distribution of junction parameters}

Before discussing the impact of a distribution of junction parameters such as RA and anisotropy constant on a distribution of WER, we briefly show basic properties of WER of STT switching without a distribution of junction parameters. Figure \ref{fig:fig1}(b) shows the typical examples of the logarithmic plot of the WER, $w$, as a function of current density. Here and hereafter the symbol $w$ stands for the WER. The solid, dotted, and dot-dashed curves indicate the results for $t_{p}$ = 5, 10, and 20 ns, respectively.  The WER suddenly drops just before the critical current density of $J_{c}=10$ MA/cm$^2$ because the magnetization can switch owing to thermal agitation even below $J_{c}$.  From a practical application point of view we are interested in the low WER regime, e.g. $w \sim 10^{-6}$. As shown in  Fig. \ref{fig:fig1}(b) the WER exponentially decreases with increase of $J$ in the low WER regime, which qualitatively agrees with the analytical expressions given in Refs. \cite{He2007,Liu2014}. Assuming that thermally distributed initial magnetization states determine the distribution of switching time for $J \gg J_{c}$, the WER is expressed as \cite{Liu2014}
\begin{align}
  \label{eq:wer1}
  w
  =
  1-\exp\left\{
  - 4 \Delta_{K}
  \exp\left[
    - \left(\frac{J}{J_{c}} -1\right)\frac{2 t_{p}}{t_{D}}
    \right]
  \right\},
\end{align}
where $t_{D}$ is a characteristic time scale for switching dynamics defined as
\begin{align}
  \label{eq:tD}
  t_{D}
  =
  \frac{(1+\alpha^{2})M_{s}}{2\alpha \gamma K}.
\end{align}
In the case of $w \ll 1$ Eq. \eqref{eq:wer1} is approximated as
\begin{equation}
  \label{eq:werHe}
  w
  =
  4 \Delta_{K}
  \exp\left[
    - \left(\frac{J}{J_{c}} -1\right)\frac{2 t_{p}}{t_{D}}
    \right].
\end{equation}
Although Eq. \eqref{eq:werHe} is valid for $J \gg J_{c}$ where thermal agitation field during the precession is neglected, numerical simulation results shown in Fig. \ref{fig:fig1}(b) implies that the current dependence of the WER takes the similar form as Eq. \eqref{eq:werHe} even in the region $J \sim J_{c}$ as long as $w \ll 1$. In Ref. \cite{Liu2014}, they analyzed the experimental data by treating $\Delta_{K}$ and $t_{D}$ as fitting parameters. Here we made a crude approximation that the effect of thermal agitation field can be taken into account by renormalizing the anisotropy constant. Introducing the renormalization coefficient $\xi$, the parameters $K$, $J_{c}$, and $t_{D}$ are renormalized as $\xi K$, $\xi J_{c}$, and $t_{D}/\xi$, respectively. The coefficient $\xi$ is determined by fitting the $t_{p}$ dependence of $J$ required to achieve $w=10^{-6}$. As pointed out in Refs. \cite{Bedau2010,Bedau2010a,Liu2014} the current density required for a certain switching probability is inversely proportional to $t_{p}$ when STT gives a dominant contribution to the switching dynamics.  In Fig. \ref{fig:fig1}(c) the simulation results of the $t_{p}$ dependence of $J$ for $w=10^{-6}$ is shown by the open circles. The simulation results are well fitted by the function $J = A/t_{p} + B$ with $A= 3.1 \times 10^{11}$ C/m$^{2}$ and $B=8.8\times 10^{10}$ A/m$^{2}$ shown by the solid curve. Since the fitting parameter $B$ corresponds to the renormalized critical current density, the renormalization coefficients are determined as $\xi=B/J_{c}=0.88$.
In terms of the renormalized parameters, the WER is expressed as
\begin{align}
  \label{eq:wer_approx}
  w
  =
  4 \xi\Delta_{K}
  \exp\left[
    - \left(\frac{J}{\xi J_{c}} -1\right)\frac{2 \xi\, t_{p}}{t_{D}}
    \right].
\end{align}
Figure \ref{fig:fig1}(d) shows the $J$ dependence of $w$ in the low WER regime for $t_{p}=10$ ns. The simulation results obtained by numerically solving Eq. \eqref{eq:fp_zeta} is plotted by the solid curve. Equation \eqref{eq:wer_approx} with $\xi =0.88$ and $\xi=1$ are plotted by the dotted and dot-dashed curves, respectively. The simulation results are well reproduced by Eq. \eqref{eq:wer_approx} with $\xi=0.88$.

%==================================================
%==================================================
\subsection{WER distribution due to RA distribution}
In this subsection we analyze impact of a distribution of RA on a distribution of WER. Let $r$ denote the value of RA and is assumed to obey a normal distribution with mean of $r_{0}$ and  standard deviation of $\sigma$,
\begin{align}
  \label{eq:PDFr}
  f(r)
  =
  \frac{1}{\sqrt{2\pi}\, \sigma}
  \exp\left\{
  -\frac{(r-r_{0})^{2}}{2\sigma^{2}}
  \right\}.
\end{align}
The PDF of $w$, which is denoted by $g(w)$, is obtained by using the change of the variable technique \cite{Hogg,Arai2021} as
\begin{align}
  \label{eq:cvt}
  g(w)
   & =
  f(r)\left|\frac{d r}{d w}\right|.
\end{align}
We first derive the analytical expression of $g(w)$ using Eqs. \eqref{eq:werHe},  \eqref{eq:PDFr}, and \eqref{eq:cvt}. Then we show that $g(w)$ can be expressed as a logarithmic normal distribution if $\log(w)$ can be approximated as a linear function of $\delta r$, where $\delta r = r -r_{0}$.

%--------------------------------------------------
%--------------------------------------------------
\subsubsection{Derivation of $g(w)$ based on the change of variable technique}
\begin{figure}[t]
  \centerline{ \includegraphics[width=\columnwidth]{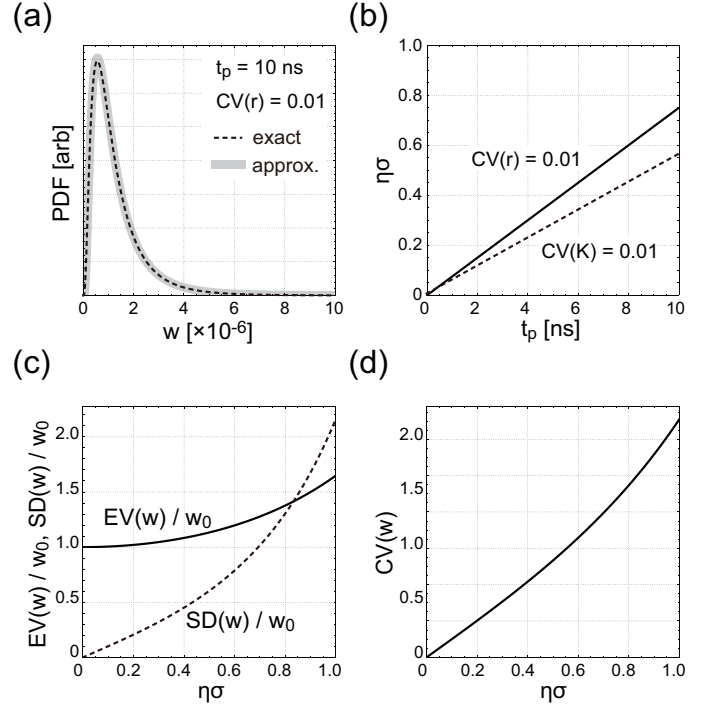}}
  \caption{
  \label{fig:fig2}
  (a) Probability density function (PDF) of write error rate (WER), $w$, for pulse width of $t_{p}$ = 10 ns. The resistance area product, $r$, is assumed to obey a normal distribution with coefficient of variation (CV) of 0.01.
  % The coefficient of variation (CV) of resistance area product, $r$, is assumed to be $CV(r)=0.01$.
  (b) $t_{p}$ dependence of $\eta\sigma$ which represents the standard deviation of $\log(w)$. Solid line represents the result for the case $r$ obeys a normal distribution with $CV(r)$ = 0.01. Dotted line represents the result for the case that the anisotropy constant, $K$, obeys a normal distribution with $CV(K)$ = 0.01.
  (c) Normalized expectation value, $EV(w)/w_{0}$, (solid) and normalized standard deviation, $S\!D(w)/w_{0}$, (dotted) as functions of $\eta\sigma$. $w_{0}$ is the WER without distribution of junction parameters.
  (d) $CV(w)$ as a function of  $\eta\sigma$.
  }
\end{figure}

Let $V$ denote the applied bias voltage defined as $J=V/r$.
The logarithm of Eq. \eqref{eq:wer_approx} is expressed as
\begin{align}
  \label{eq:logwr}
  \log(w)
  =
  -\frac{b}{r}
  +
  c,
\end{align}
where
\begin{align}
  b
  =
  \frac{\hbar\, \gamma\, V\, P\, t_{p}}{(1+\alpha^{2})M_{s}\,d\, e}
\end{align}
and
\begin{align}
  c
  =
  \log\left(4 \xi\, \Delta_{K}\right)
  +
  \frac{2 \xi\, t_{p}}{\tau_{D}}.
\end{align}
Then the derivative of $r$ in terms of $w$ is obtained as
\begin{align}
  \label{eq:drdw}
  \frac{d r}{d w}\
  =
  \frac{b}{w[c-\log(w)]^{2}}.
\end{align}
Substituting Eqs. \eqref{eq:PDFr} and \eqref{eq:drdw} into Eq. \eqref{eq:cvt} the PDF of $w$ is obtained as
\begin{align}
  \label{eq:pdfwExact}
  g(w)
  =
  \frac{q(w)^{2}}{\sqrt{2\pi}\, b\, \sigma\, w}
  \exp\left\{
  -\frac{\left[q(w)-q(w_{0})\right]^{2}}{2\sigma^{2}}
  \right\},
\end{align}
where the function $q(w)$ is defined as
\begin{align}
  q(w)=\frac{b}{c-\log(w)}.
\end{align}
In Fig. \ref{fig:fig2}(a) we plot Eq. \eqref{eq:pdfwExact} for $t_{p}$ = 10 ns by the dotted curve. The CV of $r$ is assumed to be $CV(r)=0.01$. $g(w)$ has a large skewness although $f(r)$ is assumed to be a normal distribution. In Sec. \ref{sec:approxg}, we show that $g(w)$ can be approximated as the logarithmic normal distribution shown by the thick gray curve in Fig. \ref{fig:fig2}(a).
%==============================
% Fig. 2
%==============================

%--------------------------------------------------
%--------------------------------------------------
\subsubsection{Derivation of an approximate expression of $g(w)$ using the linear approximation of $\log(w)$}
\label{sec:approxg}

Introducing
\begin{align}
  \eta=\frac{\hbar\, \gamma\, V\, P\, t_{p}}{(1+\alpha^{2})M_{s}\,d\, e \, r_{0}^{2} },
\end{align}
and take the first order of $\delta r$,  Eq. \eqref{eq:logwr} can be approximated as
\begin{align}
  \label{eq:logwdr}
  \log(w)
  =
  \log(w_{0})
  +
  \eta\, \delta r.
\end{align}
As shown in Fig. \ref{fig:fig2}(b) $\eta\sigma$ is a linear increasing function of $t_{p}$ and is less than 0.8 for $t_{p} \le $  10 ns.
Then $\delta r$ is expressed as
\begin{align}
  \label{eq:drapprox}
  \delta r
  =
  \frac{1}{\eta}\log\left(\frac{w}{w_{0}}\right),
\end{align}
of which PDF is given by
\begin{align}
  \label{eq:PDFdr}
  f(\delta r)
  =
  \frac{1}{\sqrt{2\pi}\,\sigma}
  \exp\left(-\frac{\delta r^{2}}{2\sigma^{2}}\right).
\end{align}
Substituting Eqs. \eqref{eq:drapprox} and   \eqref{eq:PDFdr} into Eq. \eqref{eq:cvt}, $g(w)$ is obtained as
\begin{align}
  \label{eq:pdfwRAapprox}
  g(w) =
  \frac{1}{\sqrt{2\pi}\, \eta \sigma  w}
  \exp\left\{
  -\frac{\left[\log(w)-\log(w_{0})\right]^{2}}{2 \eta^{2} \sigma^{2}}
  \right\},
\end{align}
which is a logarithmic normal distribution of $w$. Eq. \eqref{eq:pdfwRAapprox} tells us that $\log(w)$ obeys a normal distribution with mean of $\log(w_{0})$ and standard deviation of $\eta \sigma$.
Equation \eqref{eq:pdfwRAapprox} for $t_{p}$ = 10 ns is plotted by the thick gray curve in Fig. \ref{fig:fig2}(a), which agrees well with the exact result of Eq. \eqref{eq:pdfwExact}.

The expectation value of $w$ is given by
\begin{align}
  \label{eq:ev}
  EV(w)
  =
  \exp\left[\log(w_{0}) + \frac{\eta^{2}\sigma^{2}}{2}\right],
\end{align}
which is larger than $w_{0}$ and increases with increase of $\eta \sigma$.
The standard deviation is given by
\begin{align}
  \label{eq:sd}
  S\!D(w)
  =
  EV(w)
  \sqrt{
    \exp\left(\eta^{2}\sigma^{2}\right) -1
  } ,
\end{align}
which increases more rapidly with increase of $\eta\sigma$ compared with $EV(w)$ as shown in Fig. \ref{fig:fig2}(c) by the dotted curve.
The coefficient of variation, which is a relative measure of dispersion and is defined as the ratio of $S\!D(w)$ to $EV(w)$, is given by
\begin{align}
  \label{eq:cv}
  CV(w)
  =
  \sqrt{
    \exp\left(\eta^{2}\sigma^{2}\right) -1
  },
\end{align}
which is an increasing function of $\eta\sigma$ as shown in Fig. \ref{fig:fig2}(d). Since $\eta\sigma$ is a linear increasing function of $t_{p}$, $CV(w)$ can be reduced by decreasing $t_{p}$.

%==================================================
%==================================================
\subsection{WER distribution due to a distribution of anisotropy constant}
In this subsection we study a distribution of WER caused by a distribution of anisotropy constant, $K$. We assume that $K$ obeys a normal distribution with mean of $K_{0}$ and standard deviation of $\sigma$. The probability distribution function of $\delta K=K- K_{0}$ is given by
\begin{align}
  \label{eq:fdeltaK}
  f(\delta K)
  =
  \frac{1}{\sqrt{2\pi}\,\sigma}
  \exp\left(-\frac{\delta K ^{2}}{2\sigma^{2}}\right).
\end{align}
Similar to Eq. \eqref{eq:logwdr} we approximate $\log(w)$ up to the first order of $\delta K$ as
\begin{align}
  \label{eq:logw1dK}
  \log(w)
  =
   & \log(w_{0})
  +
  \eta\,
  \delta K,
\end{align}
where $w_{0}$ is the WER at $K_{0}$ and the coefficient $\eta$ is now defined as
\begin{align}
  \label{eq:etaK}
  \eta
  =  \frac{1}{K_{0}}\left(
  1 +\frac{2\xi\, t_{p}}{t_{D}^{0}}
  \right).
\end{align}
Here $t_{D}^{0} = (1+\alpha^{2}) M_{s}/(2\alpha \gamma K_{0})$.
The $t_{p}$ dependence of $\eta \sigma$ for $CV(K)=0.01$ is shown by the dotted line in Fig. \ref{fig:fig2}(b). $\eta \sigma$ is a linear increasing function of $t_{p}$ and is less than 0.6 for $t_{p} \leq$ 10 ns.
The PDF of $w$ is given by Eq. \eqref{eq:pdfwRAapprox} with $\eta$ defined by Eq. \eqref{eq:etaK}. The expectation value, standard deviation, and coefficient of variation are also given by the same equations as Eqs. \eqref{eq:ev}, \eqref{eq:sd}, and \eqref{eq:cv}, respectively, with $\eta$ defined by Eq. \eqref{eq:etaK}. Similar to the case with a distribution of $r$, the CV can be reduced by decreasing  $t_{p}$.

\subsection{Comparison with numerical simulations}
In the preceding subsections, we derive the analytical expressions for the PDF and statistical measures of WER and showed that the CV can be reduced by decreasing $t_{p}$ both for the case with a distribution of RA and anisotropy constant. In this subsection we perform numerical simulations based on the FP equation to confirm the validity of the analytical results.

%==============================
% Fig. 3
%==============================
\begin{figure}[t]
  \centerline{ \includegraphics[width=\columnwidth]{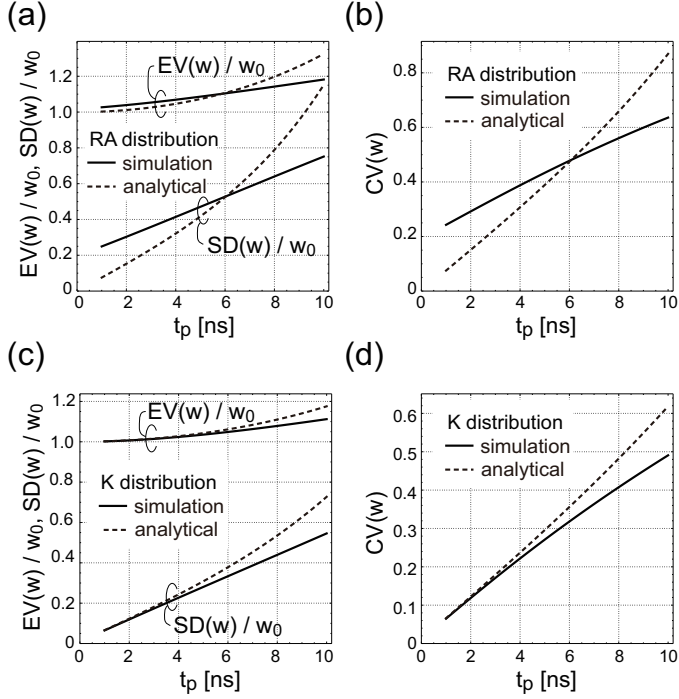}}
  \caption{
  \label{fig:fig3}
  (a) Pulse width, $t_{p}$, dependence of normalized expectation value, $EV(w)/w_{0}$, and normalized standard deviation, $S\!D(w)/w_{0}$, for the case that resistance area product (RA), $r$, obeys a normal distribution with the coefficient of variation (CV) of $CV(r)$ = 0.01. $w$ and $w_{0}$ represent the write error rate (WER) with and without a distribution of junction parameters, respectively.
  (b) $t_{p}$ dependence of the coefficient of variation of $w$, $CV(w)$, for $CV(r)$ = 0.01.
  (c) The same plot as (a) for the case that anisotropy constant, $K$, obeys a normal distribution with $CV(K)$ = 0.01.
  (d)  The same plot as (b) for $CV(K)$ = 0.01.
  In all panels the solid and dotted curves represent the simulation and  analytical results, respectively.
  }
\end{figure}

Figure \ref{fig:fig3}(a) shows the $t_{p}$ dependence of $EV(w)/w_{0}$ and $S\!D(w)/w_{0}$ for the case that RA obeys a normal distribution with $CV(r)$ = 0.01. The simulation results are represented by the solid curves, and the analytical results are plotted by the dotted curves. For both $EV(w)/w_{0}$ and $S\!D(w)/w_{0}$ the curves representing simulation results and analytical results intersect each other around $t_{p}$ = 6 ns. For $t_{p} \gtrapprox$ 6 ns, analytical results overestimate $EV(w)/w_{0}$ and $S\!D(w)/w_{0}$, and the difference between the simulation and analytical results increases with increase of $t_{p}$. The $t_{p}$ dependence of $CV(w)$ is shown in Fig. \ref{fig:fig3}(b). Both the simulation result and the analytical result are increasing function of $t_{p}$, which confirms the validity of the analytical prediction that CV can be reduced by decreasing $t_{p}$. Similar to $EV(w)$ and $S\!D(w)$, the analytical results under estimate (over estimate) the $CV(w)$ for $t_{p} \lessapprox$ 6 ns ($t_{p}\gtrapprox$ 6 ns).

The same plots for the case that $K$ obeys a normal distribution with $CV(K)$ = 0.01 are shown in Figs. \ref{fig:fig3}(c) and \ref{fig:fig3}(d). The analytical results overestimate $EV(w)$, $S\!D(w)$, and $CV(w)$ in the entire range of the plot, and the difference between the simulation results and analytical results increases with increase of $t_{p}$. Similar to the results in Fig. \ref{fig:fig3}(b), the simulation results of $CV(w)$ is an increasing function of $t_{p}$. Therefore we conclude that $CV(w)$ can be reduced by decreasing $t_{p}$ both for the case with a distribution of RA and anisotropy constant.
Simulation with $CV(K)$ = 0.01 shows that $CV(w)$ is as large as 0.49 for $t_{p}=$ 10 ns and can be reduced 0.065 by decreasing $t_{p}$ to 1 ns.

\section{Summary}
%========================================
%  Conclusion
%========================================
In summary, we theoretically study a distribution of WER of STT MRAM caused by a distribution of junction parameters, i.e. RA and anisotropy constant. Assuming that WER is much smaller than unity, and the junction parameters obey a normal distribution, we derive analytical expressions of the probability density function and statistical measures. We find that the WER obeys a logarithmic normal distribution and the CV of WER can be reduced by decreasing pulse width. The validity of the analytical results is confirmed by numerical simulations. The results provide important insights into statistical properties of STT switching and are useful for designing reliable STT MRAM.
%========================================
% Acknowledgement
%========================================
% \begin{acknowledgments}
%   This work was partly supported by JSPS KAKENHI Grants No. JP19H01108, No. JP20K12003.
% \end{acknowledgments}

%========================================
% Appendix
%========================================

\appendix
\setcounter{figure}{0}
\section{Validity of the preparation procedure of the initial distribution}
\label{sec:appendixA}
In this section we discuss the validity of the preparation procedure of the initial distribution of $\zeta$. As mentioned in the last paragraph of Sec. \ref{sec:method}, the initial distribution is prepared by relaxing $W(\zeta)$ from the delta function at $\zeta=1$ for 5 ns without applying current. In terms of the Legendre polynomials, the delta function at $\zeta=1$ is expressed as
\begin{align}
  \delta(1-\zeta)=\sum_{n=0}^{\infty}\frac{2n +1}{2}P_{n}(\zeta).
\end{align}
In numerical calculations 100 Legendre functions are used to represent the delta function.

Figure \ref{fig:figA1}(a) show the relaxation time, $t_{r}$, dependence of the expectation value of $\zeta$, which is obtained as
\begin{align}
  \langle \zeta \rangle = \int_{-1}^{1} \zeta\, W(\zeta,t_{r})\, d\zeta = \frac{2}{3}c_{1}(t_{r}).
\end{align}
The expectation value of $\zeta$ decreases with increase of $t_{r}$ and converges to the value of 0.9915 about $t_{r} =$ 1 ns.

%==============================
% Fig. A1
%==============================
\begin{figure}[t]
  \centerline{ \includegraphics[width=\columnwidth]{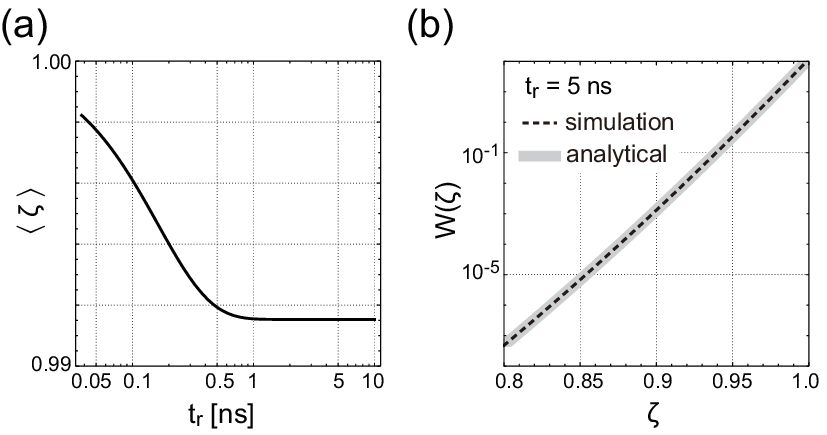}}
  \caption{
    \label{fig:figA1}
    (a) Relaxation time, $t_{r}$, dependence of the expectation value of $\zeta$, $\langle \zeta \rangle$.
    (b) Initial distribution of $\zeta$. The dotted black curve represents the initial distribution obtained by numerically solving the FP equation. The thick gray curve represents the distribution given by Eq. \eqref{eq:dist_up}.
  }
\end{figure}

In the absence of current, the system has two equivalent energy minima at $\zeta=\pm 1$ and the thermal equilibrium value of $\zeta$ is 0. However, since the thermal stability constant is assumed to be as large as 60 it takes more than tens of years to reach the thermal equilibrium. On time scale of nano-seconds the distribution of $\zeta$ is represented by the Boltzmann distribution localized on the upper hemisphere ($\zeta>0$), which is defined as
\begin{align}
  \label{eq:dist_up}
  W(\zeta)=
  \frac{2\sqrt{\Delta_{K}}}{\sqrt{\pi}\,{\rm Erfi\left(\sqrt{\Delta_{K}}\right)}}\exp\left[\Delta_{K}\zeta^{2}\right],
\end{align}
where ${\rm Erfi(\zeta)}$ denotes the imaginary error function of $\zeta$. The converged value of $\langle \zeta \rangle$ = 0.9915 is the same as the expectation value calculated using Eq. \eqref{eq:dist_up}.

In Fig. \ref{fig:figA1}(b), the distribution function, $W(\zeta)$, obtained by numerically solving the FP equation is plotted by the dotted black curve. The relaxation time is assumed to be $t_{r} = 5$ ns.
The distribution function given by Eq. \eqref{eq:dist_up} is also plotted by the thick gray curve. The good agreement between these two curves guarantees the validity of our preparation procedure of the initial distribution.

%========================================
% Bibliography
%========================================
%\bibliography{refs}

%========================================
% Appendix
%========================================

\appendix
\setcounter{figure}{0}
\section{Validity of the preparation procedure of the initial distribution}
\label{sec:appendixA}
In this section we discuss the validity of the preparation procedure of the initial distribution of $\zeta$. As mentioned in the last paragraph of Sec. \ref{sec:method}, the initial distribution is prepared by relaxing $W(\zeta)$ from the delta function at $\zeta=1$ for 5 ns without applying current. In terms of the Legendre polynomials, the delta function at $\zeta=1$ is expressed as
\begin{align}
  \delta(1-\zeta)=\sum_{n=0}^{\infty}\frac{2n +1}{2}P_{n}(\zeta).
\end{align}
In numerical calculations 100 Legendre functions are used to represent the delta function.

Figure \ref{fig:figA1}(a) show the relaxation time, $t_{r}$, dependence of the expectation value of $\zeta$, which is obtained as
\begin{align}
  \langle \zeta \rangle = \int_{-1}^{1} \zeta\, W(\zeta,t_{r})\, d\zeta = \frac{2}{3}c_{1}(t_{r}).
\end{align}
The expectation value of $\zeta$ decreases with increase of $t_{r}$ and converges to the value of 0.9915 about $t_{r} =$ 1 ns.

%==============================
% Fig. A1
%==============================
\begin{figure}[t]
  \centerline{ \includegraphics[width=\columnwidth]{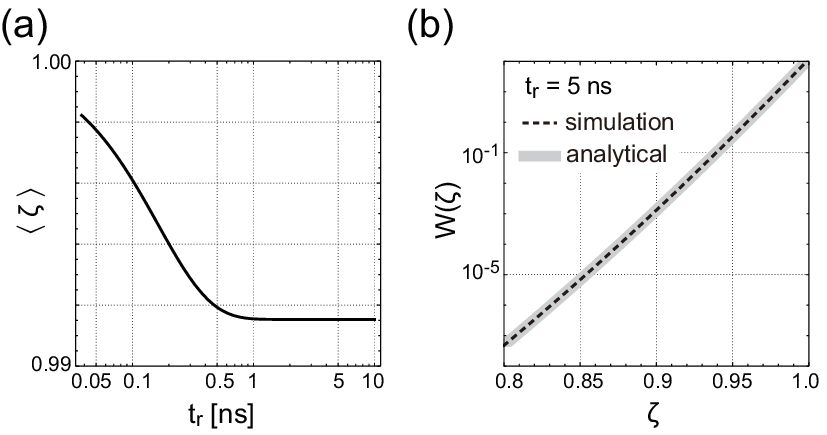}}
  \caption{
    \label{fig:figA1}
    (a) Relaxation time, $t_{r}$, dependence of the expectation value of $\zeta$, $\langle \zeta \rangle$.
    (b) Initial distribution of $\zeta$. The dotted black curve represents the initial distribution obtained by numerically solving the FP equation. The thick gray curve represents the distribution given by Eq. \eqref{eq:dist_up}.
  }
\end{figure}

In the absence of current, the system has two equivalent energy minima at $\zeta=\pm 1$ and the thermal equilibrium value of $\zeta$ is 0. However, since the thermal stability constant is assumed to be as large as 60 it takes more than tens of years to reach the thermal equilibrium. On time scale of nano-seconds the distribution of $\zeta$ is represented by the Boltzmann distribution localized on the upper hemisphere ($\zeta>0$), which is defined as
\begin{align}
  \label{eq:dist_up}
  W(\zeta)=
  \frac{2\sqrt{\Delta_{K}}}{\sqrt{\pi}\,{\rm Erfi\left(\sqrt{\Delta_{K}}\right)}}\exp\left[\Delta_{K}\zeta^{2}\right],
\end{align}
where ${\rm Erfi(\zeta)}$ denotes the imaginary error function of $\zeta$. The converged value of $\langle \zeta \rangle$ = 0.9915 is the same as the expectation value calculated using Eq. \eqref{eq:dist_up}.

In Fig. \ref{fig:figA1}(b), the distribution function, $W(\zeta)$, obtained by numerically solving the FP equation is plotted by the dotted black curve. The relaxation time is assumed to be $t_{r} = 5$ ns.
The distribution function given by Eq. \eqref{eq:dist_up} is also plotted by the thick gray curve. The good agreement between these two curves guarantees the validity of our preparation procedure of the initial distribution.

%========================================
% Bibliography
%========================================
%\bibliography{refs}

\end{document}